\documentclass[runningheads]{llncs}
\usepackage[T1]{fontenc}
\usepackage{graphicx}
\usepackage{orcidlink}
\usepackage{booktabs}
\usepackage{graphicx}
\usepackage{amsfonts}
\usepackage{amsmath} 
\usepackage{marvosym}
\usepackage[normalem]{ulem}
\useunder{\uline}{\ul}{}
\usepackage{url}
\usepackage{color}

\begin{document}
\title{TaGAT: Topology-Aware Graph Attention Network For Multi-modal Retinal Image Fusion}
%
\titlerunning{TaGAT For Multi-modal Retinal Image Fusion}
%

\author{Xin Tian\inst{1}\orcidlink{0000-0003-1168-5298}$^{(\textrm{\Letter})}$ \and
Nantheera Anantrasirichai\inst{1}\orcidlink{0000-0002-2122-5781} \and
Lindsay Nicholson\inst{2}\orcidlink{0000-0002-6942-208X} \and
Alin Achim\inst{1}\orcidlink{0000-0002-0982-7798}
}

\authorrunning{X. Tian et al.}
%
%
\institute{Visual Information Laboratory, University of Bristol, Bristol, UK \and Autoimmune Inflammation Research, University of Bristol, Bristol, UK\\
\email{\{xin.tian, n.anantrasirichai, l.nicholson, alin.achim\}@bristol.ac.uk}}

\maketitle              
\begin{abstract}
In the realm of medical image fusion, integrating information from various modalities is crucial for improving diagnostics and treatment planning, especially in retinal health, where the important features exhibit differently in different imaging modalities. Existing deep learning-based approaches insufficiently focus on retinal image fusion, and thus fail to preserve enough anatomical structure and fine vessel details in retinal image fusion. To address this, we propose the Topology-Aware Graph Attention Network (TaGAT) for multi-modal retinal image fusion, leveraging a novel Topology-Aware Encoder (TAE) with Graph Attention Networks (GAT) to effectively enhance spatial features with retinal vasculature's graph topology across modalities. The TAE encodes the base and detail features, extracted via a Long-short Range (LSR) encoder from retinal images, into the graph extracted from the retinal vessel. Within the TAE, the GAT-based Graph Information Update (GIU) block dynamically refines and aggregates the node features to generate topology-aware graph features. The updated graph features with base and detail features are combined and decoded as a fused image. Our model outperforms state-of-the-art methods in Fluorescein Fundus Angiography (FFA) with Color Fundus (CF) and Optical Coherence Tomography (OCT) with confocal microscopy retinal image fusion. The source code can be accessed via \url{https://github.com/xintian-99/TaGAT}.

\end{abstract}
\section{Introduction}

Multi-modal medical image fusion aims to combine the complementary information from various medical imaging modalities, thereby aiding in more comprehensive diagnostics and treatment planning in brain, lungs, eye/retina, and cardiac~\cite{azam2022review}. In ophthalmology, this can involve the fusion of Color Fundus (CF) images with Fluorescein Fundus Angiography (FFA), Optical Coherence Tomography (OCT) with fundus images, and OCT with confocal microscopy images~\cite{tian2019multimodal}, among others. An example illustrating the need for fusion arises when the contrast between the retinal vasculature and the background in CF is limited, thereby complicating the analysis of small retinal vessels. Conversely, FFA images enhance the visibility of the retinal vasculature by employing a fluorescent dye~\cite{hajeb2012diabetic}. The fusion of CF and FFA can integrate the high-resolution detail of pathologies in CF images with the enhanced vascular contrast from FFA. This integration furnishes a more detailed and comprehensive representation of the retinal structure~\cite{1207401}, which can facilitate the early detection, accurate diagnosis, and effective monitoring of ocular diseases such as Diabetic Retinopathy (DR)~\cite{wong2008prevalence}. The results of image fusion not only enhance the visualisation and analysis of retinal diseases by clinicians but also potentially support a range of downstream tasks, including vessel segmentation, disease classification, and monitoring of disease progression~\cite{azam2022review,ignet,cddfuse,zhao2023ddfm,zhou2019review}.

The current deep learning-based multi-modal image fusion
has achieved significant advancements with two primary branches: generation-based methods (e.g. diffusion model~\cite{ho2020denoising}, generative adversarial networks~\cite{goodfellow2014generative}), and discrimination-based methods (e.g. auto-encoder)~\cite{azam2022review}. DDFM~\cite{zhao2023ddfm} is a generative method utilising a denoising diffusion-based posterior sampling model to preserve more details for image fusion. SwinFusion~\cite{swinfusion} used cross-domain long-range learning and the Swin Transformer~\cite{liu2021swin} to efficiently integrate structure, detail, and intensity across modalities. CDDFuse~\cite{cddfuse} is an auto-encoder-based model using a decomposition loss to modulate between modality-specific and shared features extracted through a dual-branch Transformer-CNN Long-short Range (LSR) encoder to leverage CNN's proficiency in capturing local spatial details and Transformer's capability in modelling long-range dependencies~\cite{vaswani2017attention,xie2023structure}. With the advancements in feature representation capabilities of Graph Neural Networks~\cite{gcn,wu2020comprehensive}, IGNet~\cite{ignet} employed a fixed node weights GNN for cross-modality feature interaction. However, their approach to graph construction is solely based on feature space and does not incorporate the spatial structures of the images. Although these methods are effective in Visible-Infrared, MRI-CT, and MRI-PET fusion tasks, our findings reveal that they often fail to capture detailed features of the retinal vasculature and optic disc areas, particularly in abnormal retinas, when applied to retinal image fusion.

To address this gap in retinal image fusion, we introduce the Topology-Aware Graph Attention Network (TaGAT) for multi-modal retinal image fusion as shown in Fig.~\ref{method}. A Topology-Aware Encoder (TAE) is proposed to bridge the base and detail spatial features in Euclidean space with the underlying graph topology in the non-Euclidean geometric space of retinal vasculature. This leverages the consistent topological properties of vascular structures across different retinal imaging modalities, which enhances feature representation and model generalisation. The TAE utilises base and detail features extracted by an LSR encoder as node features, combined with a graph derived from the retinal vessel structure. With a Graph Attention Network (GAT), the TAE dynamically updates the graph by aggregating and refining node features, thereby connecting long-range structural features and preserving local details. Finally, a decoder is applied to reconstruct the fused image from the base, detail, and graph features. In conclusion, our contribution can be summarised as follows.
\begin{enumerate}
    \item [1)] We introduce an end-to-end framework of a topology-aware graph attention network for multi-modal retinal image fusion.
    \item [2)] We propose a GAT-based Topology-Aware Encoder, the first to bridge spatial features with the consistent graph topology of retinal vasculature across modalities. This enhances feature representation and model generalisation and ensures the preservation of important anatomical structures and fine vasculature details in the retinal image fusion.
    \item [3)] Our method achieves leading performances in retinal image fusion evaluated on both the DRFF(FFA-CF) and OCT2Confocal datasets, with exceptional preservation of fine structures, details, and textures.
\end{enumerate}
%
%
\section{Methodology}
\begin{figure}[tb]
\includegraphics[width=\textwidth]{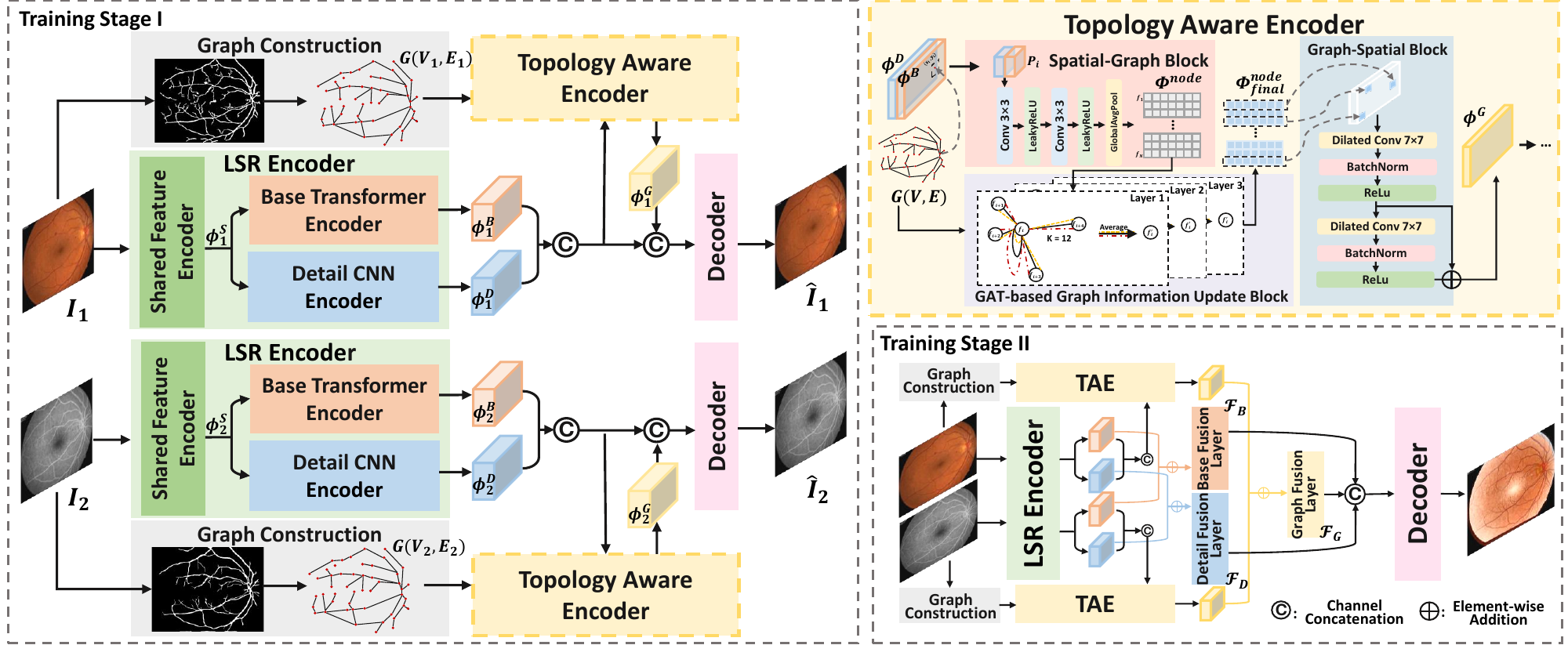}
\caption{Illustration of the proposed TaGAT framework and TAE.} \label{method}
\end{figure}

The proposed framework for multi-modal retinal image fusion is shown in Fig. \ref{method}, where the image inputs are registered, the significant features of each modality are enhanced and finally fused. We employ an LSR encoder to extract the base and detail features across modalities. The proposed GAT-based TAE encodes and updates these features with the graph topology extracted from the vessel structure. Then, the base, detail, and graph features are fused and decoded to the image domain. We employ a two-stage training strategy~\cite{cddfuse}, where the decoder in Stage I reconstructs original images and in Stage II generates fusion images.

\subsection{Graph Construction}

The graph is constructed based on a tailored wavelet-based segmentation of blood vessels~\cite{otgmreg} from retinal images \(I_{1}\) and \(I_{2}\) of Modality 1 and Modality 2, respectively. After vessel segmentation, the vascular branching points and endpoints are identified as graph nodes \(V\), \(V = \{v_i\}_{i=0}^N\), where $v_i$ is a vertex $i$ of the total $N$ vertices. These nodes \(V\) are interconnected through edges \(E\), \(E = \{e_j\}_{j=0}^M\), where $e_j$ is an edge $j$ of the total $M$ edges. The interconnection is based on the connectivity of \(V\) within the vessel network. Consequently, we have the graphs \(G_{1}(V_{1}, E_{1})\) and \(G_{2}(V_{2}, E_{2})\), which capture essential vascular characteristics such as lines, shapes, and topological configurations. By identifying the nodes' locations within the images, the graph links the graph topology (geodimensional space) and spatial domain (Euclidean space within the image domain). 

\subsection{Long-short Range Encoder}

The LSR Encoder~\cite{cddfuse} is a dual branch encoder with three components: 

i) \textit{Shared Feature Encoder} (SFE): Restormer block~\cite{restormer}-based encoder extracts shared shallow features $\Phi_{{1}}^S$ and $\Phi_{{2}}^S$ across modalities without increasing computational complexity.

ii) \textit{Base Transformer Encoder} (BTE): Lite Transformer (LT) block~\cite{DBLP:conf/iclr/WuLLLH20}-based encoder extracts low-frequency base features $\Phi_{{1}}^B$ and $\Phi_{{2}}^B$ from the shared features. 

iii) \textit{Detail CNN Encoder} (DCE): Invertible Neural Networks (INN) block~\cite{inn}-based encoder extracts high-frequency details $\Phi_{{1}}^D$ and $\Phi_{{2}}^D$ from the shared features for preserving edge and texture information in both modalities. 

\subsection{Topology Aware Encoder}
The proposed TAE encoder is designed to integrate spatial and topological information from retinal images. It comprises three main blocks:

\subsubsection{Spatial-to-Graph Block (S2G)} maps spatial features to the graph domain. The concatenated base and detail feature maps are reduced in their number of channels via convolution with a kernel size of 1 to compress features and reduce computation:
\begin{equation}
\Phi_{reduced} = \text{Conv}_{1 \times 1}\left(\text{Concat}\left(\Phi^B, \Phi^D\right)\right).
\end{equation}

Subsequently, at each node $i$ of \(G(V,E)\), the feature patch $P_i$ with the size of \(p \times p\) 
 (\(p=21)\) is extracted from $\Phi_{reduced}$ as expressed in Eq.\ref{equ:patch}, where \((x_i, y_i)\) is the spatial location of node \(i\).
\begin{equation}
    \mathbf{P}_i = \Phi_{reduced}[:, (y_i - \frac{p}{2}):(y_i + \frac{p}{2}), (x_i - \frac{p}{2}):(x_i + \frac{p}{2})].
    \label{equ:patch}
\end{equation}
Then, \(\mathbf{P}_i\) are encoded through convolutions with kernel size 3 and LeakyReLU activations.
The global average pooling is applied to yield a single feature vector \(f_i\) for each node $i$:
\begin{equation}
f_i = \text{GlobalAvgPool}\left(\text{LeakyReLU}\left(\text{Conv}_{3 \times 3}\left(\text{LeakyReLU}\left(\text{Conv}_{3 \times 3}\left(\mathbf{P}_i\right)\right)\right)\right)\right).
\end{equation}

Subsequently, the final graph node feature matrix \(\mathbf{\Phi}^{node} = \{f_1, f_2, ..., f_N\}.\) \(\mathbf{\Phi}^{node} \in \mathbb{R}^{N \times C}\), where \(N\) is the number of nodes and \(C\) is the dimensionality of the feature vectors. The \(\mathbf{\Phi}^{node}\) implicitly integrates spatial attributes into the graph for further updating by GAT-GIU block.

\subsubsection{GAT-based Graph Information Update Block (GAT-GIU)} employs a multi-layer, multi-head GAT~\cite{gat} structure to iteratively refine node features \(\mathbf{\Phi}^{node}\) through attention-driven, weighted aggregation of neighbourhood information. The node features first undergo a linear transformation, \(\mathbf{H} = \mathbf{\Phi}^{node}W\), where $W$ is a weight matrix. Then an attention mechanism is employed to compute attention coefficients $e_{ij}$ for each node pair \((i, j)\). These coefficients are then normalised across all neighbours to ensure selective attention $\alpha_{ij}$:
\begin{equation}
\alpha_{ij} = \frac{\exp(e_{ij})}{\sum_{k \in \mathcal{N}_i} \exp(e_{ik})}, \: \: \: e_{ij} = \text{LeakyReLU}(\mathbf{a}^T[\mathbf{H}_i \| \mathbf{H}_j]), \\
\end{equation}
and the updated node features \(\mathbf{\Phi}^{node}_{updated}\) based on weighted neighborhood feature aggregation is defined as Eq. \ref{equ:phiupdate}. The outputs of \(K\) heads multi-head attention are averaged to produce \(\mathbf{\Phi}^{node}_{final}\). ELU is the Exponential Linear Unit~\cite{elu}.
\begin{equation}
\mathbf{\Phi}^{node}_{final} = \frac{1}{K} \sum_{k=1}^{K} \mathbf{\Phi}^{node}_{updated, k}, \: \:  \mathbf{\Phi}^{node}_{updated} = \text{ELU}\left(\sum_{j \in \mathcal{N}_i} \alpha_{ij} \mathbf{H}_j\right).
\label{equ:phiupdate}
\end{equation}
\subsubsection{Graph-to-Spatial Block (G2S)} first maps \(\mathbf{\Phi}^{node}_{final}\) onto a feature matrix based on corresponding node coordinates and then diffuse across the spatial domain using convolutions with a larger kernel size (7 $\times$ 7) and dilation to extend the spatial influence of node features. Skip connections are applied to merge the adjusted features for producing the final enhanced feature \(\mathbf{\Phi}^{G}\) with graph topology.

\subsection{Decoder}

The decoder reduces the channel of concatenated features and uses the Restomer block for decoding. In training stage \uppercase\expandafter{\romannumeral1}, it concatenates encoders extracted features via channel dimension as input and the reconstructed image $\hat{I_{1}}$ and $\hat{I_{2}}$ as output. In the training stage \uppercase\expandafter{\romannumeral2}, the decoder takes the concatenation of features processed through feature fusion layers $\mathcal{F_B}$, $\mathcal{F_D}$, and $\mathcal{F_G}$ (base, detail, and graph, respectively) as input and a fused image \(I_f\) as the output.

\subsection{Loss Function}
For Training Stage \uppercase\expandafter{\romannumeral1}, the total loss $\mathcal{L}_{total}^{\uppercase\expandafter{\romannumeral1}}$ is:
\begin{equation}\label{equ:loss1}
    \mathcal{L}_{total}^{\uppercase\expandafter{\romannumeral1}}
    = \mathcal{L}_{1} + \alpha_1\mathcal{L}_{2} + \alpha_2\mathcal{L}_{decomp}+ \alpha_3\mathcal{L}_{graph},
\end{equation}
$\mathcal{L}_{1}$ and $\mathcal{L}_{2}$ are the reconstruction losses~\cite{cddfuse} for Modality 1 and Modality 2, ensuring original image information preservation during encoding and decoding:
\begin{equation}\label{equ:loss_ffa}
    \mathcal{L}_{m} =\mathcal{L}_{int}^{\uppercase\expandafter{\romannumeral1}}(I_{m},\hat{I_{m}}) + \mu \mathcal{L}_{SSIM}(I_{m},\hat{I_{m}}), \: \: m \in \{1,2\}
\end{equation}
where $\mathcal{L}_{int}^{\uppercase\expandafter{\romannumeral1}}\!=\!\| I_{m}-\hat{I_{m}} \|_2^2$ is the intensity loss~\cite{DBLP:journals/inffus/TangYM22}, and $\mathcal{L}_{SSIM}(I_{m},\hat{I_{m}})\!=\! 1\!-\!SSIM(I_{m},\hat I_{m})$~\cite{wang2004image}. 

The $\mathcal{L}_{decomp}$ denotes the feature decomposition loss~\cite{cddfuse} :
\begin{equation}\label{equ:loss_d}
    \mathcal{L}_{decomp} = \frac{\left(\mathcal{L}_{CC}^D\right)^2}{\mathcal{L}_{CC}^B} = \frac{\left(\mathcal{CC}\left({\Phi_1}^D,{\Phi_2}^D\right)\right)^2}{\mathcal{CC}\left({\Phi_1}^B,{\Phi_2}^B\right)+\epsilon}
\end{equation}
where $\mathcal{CC}\left(\cdot, \cdot\right)$ is the correlation coefficient operator, $\epsilon$ is set to 1.01 keeping this term positive. The $\mathcal{L}_{decomp}$ extracts detail and base features by modulating the correlation between low-frequency and high-frequency components accordingly.

\begin{equation}\label{equ:loss_g}
    \mathcal{L}_{graph} = 1 - \frac{\langle {\Phi_1}^G, {\Phi_2}^G \rangle}{| {\Phi_1}^G | \cdot | {\Phi_2}^G |}
\end{equation}
The $\mathcal{L}_{graph}$ employed cosine similarity emphasises the directional alignment of GAT-encoded features over magnitude to maintain the vascular topology similarity and adjacency information across modalities.

For Training Stage \uppercase\expandafter{\romannumeral2}, the total loss is:
\begin{equation}\label{equ:loss2}
    \mathcal{L}_{total}^{\uppercase\expandafter{\romannumeral2}}
    = \mathcal{L}_{int}^{\uppercase\expandafter{\romannumeral2}} + \alpha_3\mathcal{L}_{graph} + \alpha_4\mathcal{L}_{grad} + \alpha_5\mathcal{L}_{decomp},
\end{equation}
where $\mathcal{L}_{grad}=\frac{1}{HW}\|\left|\nabla I_f\right|-\max (\left|\nabla I_{1}\right|,\left|\nabla I_{2}\right|)\|_1$ ensures more fine-grained texture information~\cite{DBLP:journals/inffus/TangYM22}. $\mathcal{L}_{int}^{\uppercase\expandafter{\romannumeral2}}=\frac{1}{HW}\|I_f-\max (I_{1}, I_{2})\|_1$. $\nabla$ is the Sobel gradient operator. $\alpha_{1-5}$ are the hyperparameters.

\section{Experimental Results and Discussion}

\subsection{Datasets}
Two datasets were involved in our experiments.
i) \textbf{DRFF}~\cite{hajeb2012diabetic}: The DRFF dataset comprises 30 abnormal and 29 normal unregistered FFA-CF pairs. We applied the segmentation and registration method from~\cite{otgmreg} and used a subset comprising 20 normal and 20 abnormal pairs for training and 19 pairs for testing. Data augmentation with flipping, rotating by ±8 degrees, and translating by ±20 pixels are applied.
ii) \textbf{OCT2Confocal}~\cite{tian2023oct2confocal}: The dataset has paired grayscale OCT and corresponding coloured confocal microscopy retinal images from 3 mice afflicted with autoimmune uveitis. The registration is through manual registration and confirmed by an ophthalmologist. We use this data to test models trained on the DRFF dataset.
\subsection{Experimental Setup and Evaluation Metrics}
Our computational experiments were conducted on a high-performance computing environment featuring NVIDIA Tesla V100 GPUs (32 GB). Training is conducted with the first stage 40 epochs and the second stage 80 epochs. The images are resized to 288$\times$360 pixels for training with a batch size of 1 due to memory limitation. The Adam optimiser is employed, with an initial learning rate of $10^{-4}$ and decay by a factor of 0.5 every 20 epochs.

In the LSR encoder, both the SFE and BTE are configured with 4 Restormer blocks, utilising 4 attention heads (compared to 8 as used in CDDFuse~\cite{cddfuse}) within a 64-dimensional embedding space. The GIU block is equipped with 12 attention heads, each operating in a 64-dimensional space. The decoder employs 4 Restormer blocks with 4 attention heads. For the loss function, the weighting coefficients $\alpha_1$ through $\alpha_5$ are finely tuned to the values of 1, 2, 0.5, 10, and 2. 

We use eight metrics to measure the fusion results \cite{ma2019infrared}: entropy (EN), standard deviation (SD), spatial frequency (SF), mutual information (MI), sum of the correlations of differences (SCD), visual information fidelity (VIF), $Q^{AB/F}$ and SSIM. Higher metrics indicate that a fusion image is better. 
\begin{figure}[tb]
\includegraphics[width=\textwidth]{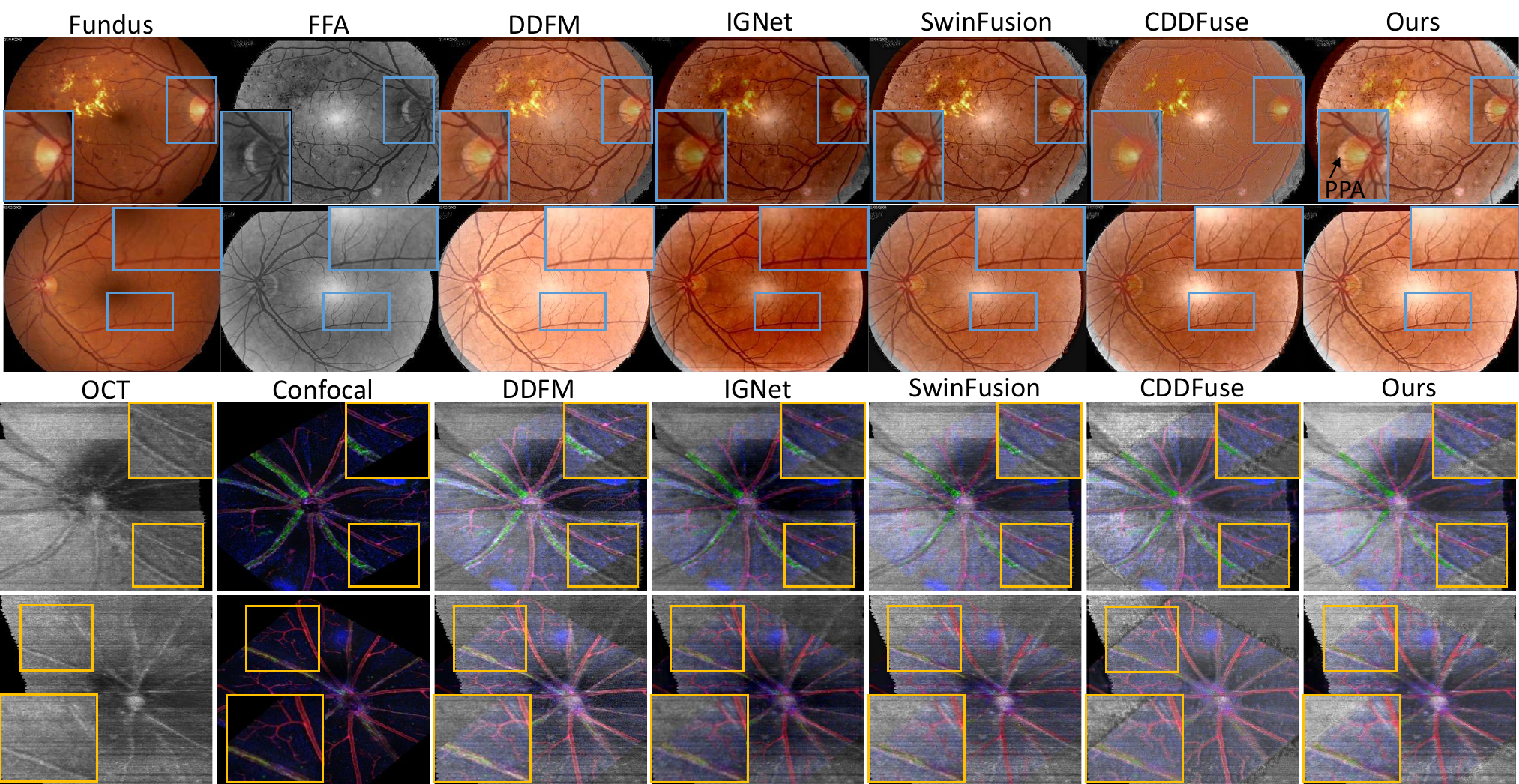}
\caption{Visual comparison results in DRFF and OCT2Confocal.} \label{all}
\end{figure}

\subsection{Benchmarking Results}
We tested our model and compared the fusion results with the existing benchmarks including DDFM~\cite{zhao2023ddfm}, IGNet~\cite{ignet}, SwinFusion~\cite{swinfusion}, and CDDFuse~\cite{cddfuse}.

For the FFA-CF dataset, Table~\ref{tab:alldataresult} (left) showcases our model's leading performance, particularly highlighted by the highest SD and SF, indicating marked improvements in detail and structure preservation. Despite competitive VIF and SSIM scores from SwinFusion and CDDFuse, our model demonstrates balanced performance across all metrics. Additionally, visual results in Fig.~\ref{all} (top 3 rows) highlight our model's ability to clearly delineate the optic disc's shape and maintain fine vasculature and texture details, indicating the TAE effectively encodes vessel topology into features to enhance focus on vasculature. The first row in Fig.~\ref{all} also demonstrates our model's ability to reveal the Peripapillary Atrophy (PPA) region, characterised by atrophic changes and irregular retinal pigmentation around the optic disc, which is challenging to discern in standard CF. Identifying PPA is crucial for diagnosing conditions like diabetic retinopathy.


For OCT2Confocal dataset, our method outperforms other methods as shown in Table~\ref{tab:alldataresult} (right). However, visual results in Fig.~\ref{all} (bottom 2 rows) indicate none of the tested methods offers sufficient detail and clarity preservation across both modalities. Compared to CDDFuse, our results show a denoising effect and exhibit fewer border artefacts. This improvement is attributed to our topology-aware graph feature, which emphasises vessel-related information for fusion, effectively minimising irrelevant features such as noise.


\begin{table}[tb]
\centering
\caption{Quantitative results of the DRFF and OCT2Confocal retinal image fusion. \textbf{Bold} and {\ul Underline} show the best and second-best results, respectively.}
\label{tab:alldataresult}
\resizebox{\textwidth}{!}{%
\begin{tabular}{@{}l|ccccccccc|cccccccc@{}}
\toprule
\multicolumn{1}{c|}{} & \multicolumn{8}{c}{DRFF Retinal Images} &  & \multicolumn{8}{c}{OCT2Confocal Retinal Images} \\ \cmidrule{2-18} 
\multicolumn{1}{c|}{} & EN & SD & SF & MI & SCD & VIF & $Q^{AB/F}$ & SSIM &  & EN & SD & SF & MI & SCD & VIF & $Q^{AB/F}$ & SSIM \\
\midrule
DDFM~\cite{zhao2023ddfm} & 6.55 & {\ul 58.9} & 14.07 & 1.41 & {\ul 1.3} & 0.22 & 0.21 & 0.27 &  & 7.01 & {\ul 38.21} & 15.98 & 1.08 & {\ul 1.4} & 0.17 & 0.19 & 0.36 \\
IGNet~\cite{ignet} & 6.75 & 39.12 & 12.28 & 1.61 & 0.29 & 0.66 & 0.49 & 0.91 &  & 3.45 & 16.16 & 5.41 & 0.6 & 0.54 & 0.19 & 0.17 & 0.36 \\
SwinFusion~\cite{swinfusion} & 6.86 & 49.01 & 16.41 & {\ul 3.15} & 0.66 & 1.03 & {\ul 0.65} & {\ul 0.99} &  & \textbf{7.12} & 41.05 & 17.2 & {\ul 2.71} & 1.19 & {\ul 0.73} & {\ul 0.6} & {\ul 0.95} \\
CDDFuse~\cite{cddfuse} & \textbf{7.08} & 57.22 & {\ul 17.06} & 2.88 & 0.72 & {\ul 0.91} & 0.64 & \textbf{1.01} &  & {\ul 7.05} & 41.28 & {\ul 17.87} & 1.56 & 1.28 & 0.48 & 0.4 & 0.87 \\
\textbf{Ours} & {\ul 6.97} & \textbf{69} & \textbf{19.22} & \textbf{3.45} & \textbf{1.52} & \textbf{1.03} & \textbf{0.66} & 0.96 &  & 6.94 & \textbf{67.59} & \textbf{19.18} & \textbf{3.36} & \textbf{1.54} & \textbf{1} & \textbf{0.66} & \textbf{0.98} \\ \bottomrule
\end{tabular}%
}
\end{table}

\begin{table}[tb]
\centering
\caption{Ablation experiments results with DRFF. \textbf{Bold} indicates the best value.}
\label{tab:ablation}
{%
\begin{tabular}{@{}cccccc@{}}
\toprule
    & Configurations   & SD    & MI   & VIF  & SSIM \\ \midrule
I   & w/o ${L}_{graph}$            & 67.23 & 3.4  & 1.01 & 0.96 \\
II  & w/o G2S   & 65    & 2.61 & 0.66 & 0.84 \\
III & w/o ${\Phi}^B$ and ${\Phi}^D$ for Decoder      & 39.26 & 1.02 & 0.25 & 0.41 \\
IV & w/o ${\Phi}^G$      & 66.69 & 3.38 & 0.64 & 0.92 \\
V  & GAT → GCN        & 68.29 & 2.94 & 0.88 & 0.95 \\ \midrule
\multicolumn{2}{c}{\textbf{Ours}} & \textbf{69} & \textbf{3.45} & \textbf{1.03} & \textbf{0.96} \\ \bottomrule
\end{tabular}%
}
\end{table}

\subsubsection{Ablation Studies}
We verified the effectiveness of i) the graph loss ${L}_{graph}$, ii) G2S block, iii) ${\Phi}^B$ and ${\Phi}^D$, iv) ${\Phi}^G$, and v) GAT. Table~\ref{tab:ablation} shows that without \({L}_{graph}\) or G2S block leads to a slight decrease in all metrics highlighting their role in refining the feature representations. Removing \({\Phi}^G\) slightly diminishes the performance due to less attention around vessels. When excluding \({\Phi}^B\) and \({\Phi}^D\) from the decoder, the marked reduction across all evaluated metrics suggests that graph-related features are insufficient for reconstructing a full Euclidean space image primarily due to lack of the detailed pixel-level information. Substituting GAT with normal GCN~\cite{gcn} is to validate the utility of dynamic attention mechanisms. The reduced performance suggests the effectiveness of the dynamic attention mechanisms for GAT in feature aggregation. 

\section{Conclusion}
This paper presents a multimodal retinal image fusion method with a novel GAT-based TAE feature encoder that effectively bridges spatial-temporal and graph topology characteristics across different modalities. Our approach has demonstrated superior performance in enhancing key feature visualisation such as the clarity of optic disc and PPA, the preservation of fine vasculature and texture details in FFA-CF fusion with the ablation studies validating the significance of each model component. In future work, we aim to enhance the fusion of low-quality and high-resolution images and extend our approach to other types of medical images with vessel structure, such as brain MRI and CT scans.

\begin{credits}
\subsubsection{\ackname} Xin is supported by the China Scholarship Council.

\subsubsection{\discintname}
The authors have no competing interests to declare that are relevant to the content of this article.
\end{credits}
%
%
%
\bibliographystyle{splncs04}
\bibliography{Paper-0115}

\begin{thebibliography}{10}
\providecommand{\url}[1]{\texttt{#1}}
\providecommand{\urlprefix}{URL }
\providecommand{\doi}[1]{https://doi.org/#1}

\bibitem{azam2022review}
Azam, M.A., Khan, K.B., Salahuddin, S., Rehman, E., Khan, S.A., Khan, M.A., Kadry, S., Gandomi, A.H.: A review on multimodal medical image fusion: Compendious analysis of medical modalities, multimodal databases, fusion techniques and quality metrics. Computers in biology and medicine  \textbf{144},  105253 (2022)

\bibitem{elu}
Clevert, D.A., Unterthiner, T., Hochreiter, S.: Fast and accurate deep network learning by exponential linear units (elus). arXiv preprint arXiv:1511.07289  (2015)

\bibitem{inn}
Dinh, L., Sohl{-}Dickstein, J., Bengio, S.: Density estimation using real {NVP}. In: {ICLR} (2017)

\bibitem{goodfellow2014generative}
Goodfellow, I., Pouget-Abadie, J., Mirza, M., Xu, B., Warde-Farley, D., Ozair, S., Courville, A., Bengio, Y.: Generative adversarial nets. Advances in neural information processing systems  \textbf{27} (2014)

\bibitem{hajeb2012diabetic}
Hajeb Mohammad~Alipour, S., Rabbani, H., Akhlaghi, M.R., et~al.: Diabetic retinopathy grading by digital curvelet transform. Computational and mathematical methods in medicine  \textbf{2012} (2012)

\bibitem{ho2020denoising}
Ho, J., Jain, A., Abbeel, P.: Denoising diffusion probabilistic models. Advances in neural information processing systems  \textbf{33},  6840--6851 (2020)

\bibitem{gcn}
Kipf, T.N., Welling, M.: Semi-supervised classification with graph convolutional networks. arXiv preprint arXiv:1609.02907  (2016)

\bibitem{1207401}
Laliberte, F., Gagnon, L., Sheng, Y.: Registration and fusion of retinal images-an evaluation study. IEEE Transactions on Medical Imaging  \textbf{22}(5),  661--673 (2003)

\bibitem{ignet}
Li, J., Chen, J., Liu, J., Ma, H.: Learning a graph neural network with cross modality interaction for image fusion. In: Proceedings of the 31st ACM International Conference on Multimedia. pp. 4471--4479 (2023)

\bibitem{liu2021swin}
Liu, Z., Lin, Y., Cao, Y., Hu, H., Wei, Y., Zhang, Z., Lin, S., Guo, B.: Swin transformer: Hierarchical vision transformer using shifted windows. In: Proceedings of the IEEE/CVF international conference on computer vision. pp. 10012--10022 (2021)

\bibitem{ma2019infrared}
Ma, J., Ma, Y., Li, C.: Infrared and visible image fusion methods and applications: A survey. Information Fusion  \textbf{45},  153--178 (2019)

\bibitem{swinfusion}
Ma, J., Tang, L., Fan, F., Huang, J., Mei, X., Ma, Y.: Swinfusion: Cross-domain long-range learning for general image fusion via swin transformer. IEEE/CAA Journal of Automatica Sinica  \textbf{9}(7),  1200--1217 (2022)

\bibitem{DBLP:journals/inffus/TangYM22}
Tang, L., Yuan, J., Ma, J.: Image fusion in the loop of high-level vision tasks: {A} semantic-aware real-time infrared and visible image fusion network. Inf. Fusion  \textbf{82},  28--42 (2022)

\bibitem{otgmreg}
Tian, X., Anantrasirichai, N., Nicholson, L., Achim, A.: Optimal transport-based graph matching for {3D} retinal {OCT} image registration. In: 2022 IEEE International Conference on Image Processing (ICIP). pp. 2791--2795. IEEE (2022)

\bibitem{tian2023oct2confocal}
Tian, X., Anantrasirichai, N., Nicholson, L., Achim, A.: {OCT2Confocal}: {3D} cyclegan based translation of retinal {OCT} images to confocal microscopy. arXiv preprint arXiv:2311.10902  (2023)

\bibitem{tian2019multimodal}
Tian, X., Zheng, R., Chu, C.J., Bell, O.H., Nicholson, L.B., Achim, A.: Multimodal retinal image registration and fusion based on sparse regularization via a generalized minimax-concave penalty. In: ICASSP 2019-2019 IEEE International Conference on Acoustics, Speech and Signal Processing (ICASSP). pp. 1010--1014. IEEE (2019)

\bibitem{vaswani2017attention}
Vaswani, A., Shazeer, N., Parmar, N., Uszkoreit, J., Jones, L., Gomez, A.N., Kaiser, {\L}., Polosukhin, I.: Attention is all you need. Advances in neural information processing systems  \textbf{30} (2017)

\bibitem{gat}
Velickovic, P., Cucurull, G., Casanova, A., Romero, A., Lio, P., Bengio, Y., et~al.: Graph attention networks. stat  \textbf{1050}(20),  10--48550 (2017)

\bibitem{wang2004image}
Wang, Z., Bovik, A.C., Sheikh, H.R., Simoncelli, E.P., et~al.: Image quality assessment: from error visibility to structural similarity. IEEE TIP  \textbf{13}(4),  600--612 (2004)

\bibitem{wong2008prevalence}
Wong, T.Y., Cheung, N., Tay, W.T., Wang, J.J., Aung, T., Saw, S.M., Lim, S.C., Tai, E.S., Mitchell, P.: Prevalence and risk factors for diabetic retinopathy: the singapore malay eye study. Ophthalmology  \textbf{115}(11),  1869--1875 (2008)

\bibitem{DBLP:conf/iclr/WuLLLH20}
Wu, Z., Liu, Z., Lin, J., Lin, Y., Han, S.: Lite transformer with long-short range attention. In: {ICLR} (2020)

\bibitem{wu2020comprehensive}
Wu, Z., Pan, S., Chen, F., Long, G., Zhang, C., Philip, S.Y.: A comprehensive survey on graph neural networks. IEEE transactions on neural networks and learning systems  \textbf{32}(1),  4--24 (2020)

\bibitem{xie2023structure}
Xie, H., Huang, Z., Leung, F.H., Ju, Y., Zheng, Y.P., Ling, S.H.: A structure-affinity dual attention-based network to segment spine for scoliosis assessment. In: 2023 IEEE International Conference on Bioinformatics and Biomedicine (BIBM). pp. 1567--1574. IEEE (2023)

\bibitem{restormer}
Zamir, S.W., Arora, A., Khan, S., Hayat, M., Khan, F.S., Yang, M.: {Restormer}: {E}fficient transformer for high-resolution image restoration. In: {CVPR}. pp. 5718--5729 (2022)

\bibitem{cddfuse}
Zhao, Z., Bai, H., Zhang, J., Zhang, Y., Xu, S., Lin, Z., Timofte, R., Van~Gool, L.: {CDDFuse}: {C}orrelation-driven dual-branch feature decomposition for multi-modality image fusion. In: Proceedings of the IEEE/CVF Conference on Computer Vision and Pattern Recognition (CVPR). pp. 5906--5916 (June 2023)

\bibitem{zhao2023ddfm}
Zhao, Z., Bai, H., Zhu, Y., Zhang, J., Xu, S., Zhang, Y., Zhang, K., Meng, D., Timofte, R., Van~Gool, L.: {DDFM}: denoising diffusion model for multi-modality image fusion. arXiv preprint arXiv:2303.06840  (2023)

\bibitem{zhou2019review}
Zhou, T., Ruan, S., Canu, S.: A review: Deep learning for medical image segmentation using multi-modality fusion. Array  \textbf{3},  100004 (2019)

\end{thebibliography}
%




\end{document}